\documentclass[smallextended]{svjour3}
\smartqed
  \usepackage{color}
  \usepackage{newlfont}
  \usepackage{graphicx}
  \usepackage{amssymb}
  \usepackage{amsmath}
  \usepackage{bm}
  \usepackage{verbatim}
  \usepackage[latin1]{inputenc}
  \usepackage{bbm}
  \usepackage{mathptmx}
\newcommand{\ket}[1]{\mbox{$ | #1 \rangle $}}
\newcommand{\bra}[1]{\mbox{$ \langle #1 | $}}
\newcommand{\be}{\begin{equation}}
\newcommand{\ee}{\end{equation}}
\newcommand{\beq}{\begin{eqnarray}}
\newcommand{\eeq}{\end{eqnarray}}
\journalname{Quantum Information Processing}

\begin{document}

\title{Weak quantum discord}
\author{P. R. Dieguez \and R. M. Angelo}
\institute{P. R. Dieguez \at
           Department of Physics, Federal University of Paran\'a, 
           P.O. Box 19044, 81531-980 Curitiba, Paran\'a, Brazil
           \email{dieguez.pr@gmail.com}
           \and 
           R. M. Angelo \at 
           Department of Physics, Federal University of Paran\'a, 
           P.O. Box 19044, 81531-980 Curitiba, Paran\'a, Brazil
           Tel.: +55 41 33613092\\
           Fax: +55 41 33613418\\
           \email{renato@fisica.ufpr.br}}
           
\date{Received: date / Accepted: date}

\maketitle

\begin{abstract} 
Originally introduced as the difference between two possible forms of quantum mutual information, quantum discord has posteriorly been shown to admit a formulation according to which it measures a distance between the state under scrutiny and the closest projectively measured (non-discordant) state. Recently, it has been shown that quantum discord results in higher values when projective measurements are substituted by weak measurements. This sounds paradoxical since weaker measurements should imply weaker disturbance and, thus, a smaller distance. In this work we solve this puzzle by presenting a quantifier and an underlying interpretation for what we call weak quantum discord. As a by-product, we introduce the notion of symmetrical weak quantum discord.
% \PACS{PACS 03.67.Mn \and PACS 03.65.Ud \and PACS 03.67.-a}
\end{abstract}

%03.67.-a	Quantum information
%03.67.Mn	Entanglement measures, witnesses, and other characterizations
%03.65.Ud	Entanglement and quantum nonlocality (e.g. EPR paradox, Bell's inequalities, GHZ states, etc.) (for entanglement production and manipulation, see 03.67.Bg; for entanglement measures, witnesses etc., see 03.67.Mn; for entanglement in Bose-Einstein condensates, see 03.75.Gg)

%=================================================
\section{Introduction}
\label{intro}

Soon after its inception, quantum mechanics was claimed, by Einstein, Podolsky, and Rosen (EPR) \cite{EPR35}, not to be a complete theory. The argument put forward by EPR made use of a kind of state whose correlations would reveal themselves to be ``{\em the} characteristic trait of quantum mechanics'', as seminally pondered by Schrödinger~\cite{schrodinger35}. These correlations---then called entanglement---were posteriorly shown to be necessary elements for the violation of Bell's hypothesis of local causality \cite{bell64}. Modernly defined as a class of correlations that cannot be created via local operations and classical communication~\cite{h309}, entanglement is by now widely accepted as a fundamental resource for quantum technology \cite{vincenzo00,nielsen00,galindo02,buhrman10} and an important mechanism in foundational approaches \cite{popescu94,zurek03,angelo15,bilobran15,dieguez18}.

However, as history has shown, entanglement is by no means the last word on quantum correlations. In 2001, Ollivier and Zurek \cite{ollivier01}, and Henderson and Vedral \cite{henderson01} independently, discovered a type of quantum correlation that can occur even for non-entangled states. These correlations are captured by a quantifier called {\em quantum discord}, which will be the focus of the present work (see Ref. \cite{celeri11} for a review of the remarkable developments associated with quantum discord). Other quantumness quantifiers also gained attention in the last decades, as for instance the EPR-steering \cite{wiseman07,cavalcanti09,costa16RC}, the geometrical quantum discord \cite{dakic10}, the symmetric quantum discord \cite{rulli11} and further generalizations \cite{rossignoli10,costa13}, the Bell nonlocality \cite{maudlin92,brassard99,steiner00,kas00,bacon03,branciard11,las14,fonseca15,costa16}, and, more recently, the realism-based nonlocality \cite{gomes18}. The existence of a given hierarchy underlying many of these quantifiers \cite{costa16,gomes18} can be viewed as a theoretical evidence that the measured quantum correlations have different natures. An interesting step toward an unifying approach for several quantum correlations measures (including quantum discord) was given by Modi {\em et al} in Ref. \cite{modi10}. In this work, the authors show how to state a given quantum correlation quantifier as a ``distance'' (in terms of some entropic metric) between the state under scrutiny and a state that has been projectively measured and, therefore, does not have the corresponding quantum correlation. 

The question then naturally arises as to whether one can obtain further information about quantum correlations by using weak measurements \cite{aharonov88,aharonov14,oreshkov05} instead of the projective ones. Since a weak measurement implies a weak disturbance on the state, the entropic distance to the undisturbed state should presumably be small. It follows from this rationale that the weak-measurement induced quantum discord should be never greater than its traditional formulation. This was indeed confirmed by Li {\em et al} \cite{li16}, who employed the Hilbert-Schmidt norm to compute a weak-measurement induced geometrical quantum discord. Surprisingly, though, by introducing weak measurements in the original procedure for the derivation of the entropic quantum discord, Singh and Pati obtained what they called a {\em super quantum discord} \cite{singh14}, a quantifier that is greater than quantum discord. This fact was corroborated by a number of works via explicit calculations involving two-qubit states \cite{wang14,hu15,li15,jing17}. However, as pointed out by Xiang and Jing, who also noticed the discrepancy between the super quantum discord and the weak geometrical quantum discord in contexts involving non-inertial reference frames \cite{xiang14}, there seems to be some inconsistency in all this.

The present work aims at solving this puzzle by introducing a formulation for what we call {\em weak quantum discord}. In Sec. \ref{Sec:RQD} we revisit several definitions of quantum discord and set up the room for the presentation of our main discussion. In Sec. \ref{Sec:WQD} we show how to consistently introduce the weak quantum discord and then prove that it is never greater than the quantum discord. In particular, it is shown that the weak quantum discord goes to zero with the intensity of the measurement. The meaning of the introduced measure is discussed in Sec. \ref{meaning} and a case study is presented in Sec. \ref{example}. As a by-product of our approach, we introduce in Sec. \ref{Sec:SyWQD} the {\em symmetrical weak quantum discord} and compare the aforementioned quantifiers via the concepts of hierarchy and ordering of quantum correlations. We close this work in Sec. \ref{Sec:C} with our conclusions and perspectives.

%===================================
\section{Revisiting Quantum Discord}
\label{Sec:RQD}

Quantum discord (QD) originally appeared as the breakdown, at the quantum level, of a given equivalence in the classical information theory. Consider two random variables $X$ and $Y$, for which the joint probability distribution of getting outcomes $x$ and $y$, respectively, is $p_{x,y}$. The Shannon entropy $H(X,Y)=-\sum_{x,y}p_{x,y}\ln{p_{x,y}}$ quantifies the ignorance that an observer has about these random variables. On the other hand, $H(X)=-\sum_xp_x\ln{p_x}$ and $H(Y)=-\sum_yp_y\ln{p_y}$ quantify the amount of ignorance specifically associated with each variable, where $p_{x(y)}=\sum_{y(x)}p_{x,y}$ denotes the marginal probability distribution associated with the variable $X(Y)$. The classical notion of mutual information, which is formally written as 
\be
I_{X:Y}=H(X)+H(Y)-H(X,Y),
\label{Icl}
\ee 
encapsulates the amount of information about $Y$ that is codified in $X$, and vice-versa. In this capacity, mutual information is a measure of correlations. Interestingly, there is another formula for the mutual information which makes explicit reference to the measurement process:
\be 
J_{X:Y}=H(X)-\sum_yp_yH(X|y),
\label{Jcl}
\ee 
where $H(X|y)=-\sum_xp_{x|y}\ln{p_{x|y}}$ is the entropy of $X$ conditioned to the outcome $y$ and $H(X|Y)=\sum_yp_yH(X|y)$ is the (average) conditional entropy. Now, using the very definition of conditional probability, $p_{x|y}=p_{x,y}/p_y$, one may verify that $J_{X:Y}=I_{X:Y}$.

In 2001, Ollivier and Zurek \cite{ollivier01} noted that such equivalence cannot be established in the quantum domain. On the one hand, the quantum counterpart of the mutual information \eqref{Icl} can be directly written as 
\be 
I(\rho)=S(\rho_{\cal{A}})+S(\rho_\cal{B})-S(\rho),
\label{Iq}
\ee 
where $S(\rho)=-\text{Tr}(\rho\ln{\rho})$ is the von Neumann entropy, $\rho$ is a density operator acting on the composite space $\cal{H_A\otimes H_B}$, and $\rho_{\cal{A(B)}}=\text{Tr}_{\cal{B(A)}}\rho$ is the reduced state acting on the subspace $\cal{H_{A(B)}}$. On the other hand, to devise the counterpart of the formula \eqref{Jcl} one needs to specify measurement operators and then the pertinent conditional entropy. Ollivier and Zurek proposed to use the set $\{B_b\}$ of projectors of an observable $B=\sum_bbB_b$ acting on $\cal{H_B}$. The second form of the mutual information was then proposed to be
\be 
J(\rho)=S(\rho_{\cal{A}})-\sum_bp_bS(\rho_{\cal{A}|b}),
\label{Jq}
\ee 
where $\rho_{\cal{A}|b}=\text{Tr}_{\cal{B}}[(\mathbbm{1}\otimes B_b)\rho(\mathbbm{1}\otimes B_b)]/p_b$, and $p_b=\text{Tr}[(\mathbbm{1}\otimes B_b)\rho(\mathbbm{1}\otimes B_b)]$. The second term on the right-hand side of Eq. \eqref{Jq} is the quantum counterpart of the condition entropy $H(X|Y)$. Now the crux comes. The forms \eqref{Iq} and \eqref{Jq} are not equivalent and the minimum deviation $I(\rho)-J(\rho)$ defines the so-called QD:
\be 
D_{\cal{B}}(\rho):=\min_B\Big[ \sum_bp_bS(\rho_{\cal{A}|b})+S(\rho_{\cal{B}})-S(\rho)\Big].
\label{QD}
\ee
It is clear that the QD is, by construction, a measure of quantum correlations.

By the same year, introducing the notion of classically accessible correlations, $\cal{C}(\rho)=\max_BJ(\rho)$, Henderson and Vedral \cite{henderson01} observed that one can write Eq. \eqref{QD} as $I(\rho)=D_{\cal{B}}(\rho)+\cal{C}(\rho)$, a form that allows us to interpret mutual information as the sum of purely quantum and purely classical correlations. Later on, Rulli and Sarandy \cite{rulli11} gave to QD and alternative shape. Taking the completely positive trace-preserving map
\be 
\Phi_B(\rho):=\sum_b(\mathbbm{1}\otimes B_b)\rho(\mathbbm{1}\otimes B_b)=\sum_bp_b\rho_{\cal{A}|b}\otimes B_b
\label{PhiB}
\ee 
and the identity $S(\Phi_B(\rho))=S(\Phi_B(\rho_{\cal{B}}))+\sum_bp_bS(\rho_{\cal{A}|b})$ (see the joint-entropy theorem \cite{nielsen00}), those authors wrote the QD as
\be 
D_{\cal{B}}(\rho)=\min_B\Big[I(\rho)-I(\Phi_B(\rho))\Big].
\label{QDRS}
\ee 
Besides allowing for the generalization of the notion of QD to multipartite states in a symmetrical way, which was the main goal of Rulli and Sarandy, this form admits an interesting interpretation for QD. To see this we first note that $\Phi_B(\rho)$ can be viewed as a state that has undergone an {\em unrevealed projective measurement} of the observable $B$ (see \cite{bilobran15,dieguez18} for more details). It follows that QD is the the minimum ``distance'' between $\rho$ and the projectively disturbed non-discordant state $\Phi_B(\rho)$, where the ``metric'' used is the mutual information. This is in conceptual agreement with the unified view discussed in Ref. \cite{modi10}. Of course, other metrics can be (and have been) used, including geometric ones \cite{dakic10,costa13}. 

Singh and Pati \cite{singh14} investigated what happens with quantum discord as the projective measurements are replaced with weak measurements. To this end, they employed the weak-measurement dichotomic operators introduced by Oreshkov and Brun \cite{oreshkov05}, namely, 
\be 
P_{\pm}(x)=\sqrt{\frac{1\mp \tanh x}{2}}\Pi_0  +\sqrt{\frac{1\pm \tanh x}{2}}\Pi_1, 
\label{Ppm}
\ee
with $x\in \mathbb{R}$ and $\Pi_0+\Pi_1=P^2_{+}+P^2_{-}=\mathbbm{1}$ for projectors $\Pi_0$ and $\Pi_1$ acting on $\cal{H_B}$. Singh and Pati then used these operators to construct the post-measurement state $\rho _{\cal{A}|P_{\pm}}=\text{Tr}_{\cal{B}}[(\mathbbm{1}\otimes P_{\pm})\rho(\mathbbm{1}\otimes P_{\pm})]/p_{\pm}$, with probabilities $p_{\pm}=\text{Tr}[(\mathbbm{1}\otimes P_{\pm})\rho(\mathbbm{1}\otimes P_{\pm})]$, and the ``weak conditional entropy'' $S_{x}(\cal{A}|\{P_{\pm}\})=p_+S(\rho_{\cal{A}|P_+})+p_-S(\rho _{\cal{A}|P_-})$. With that, they introduced the {\em super quantum discord} (SQD)
\be
D^x_{\cal{B}}(\rho)=\min_{\{P_{\pm}\}} \sum_{s=\pm}p_sS(\rho_{\cal{A}|P_s})+S(\rho_{\cal{B}})-S(\rho),
\label{SQD}
\ee
which is a function of $x$. The name indeed is appropriate as Singh and Pati have proved that $D_{\cal{B}}^x(\rho)\geqslant D_{\cal{B}}(\rho)$. Although the formula \eqref{SQD} for the SQD is a natural generalization of the expression \eqref{QD} for the QD, it produces a conflict with the intuition deriving from the alternative form \eqref{QDRS}: A weak measurement should imply a weak disturbance on the measured state and, therefore, a weak discord instead of a super discord. In particular, for $x=0$ we have $P_{\pm}(0)=\mathbbm{1}/\sqrt{2}$, which should imply no change in the state. Still, from the formula \eqref{SQD} we obtain $D_{\cal{B}}^{x=0}(\rho)=I(\rho)$, which is clearly non-zero. In what follows, we propose a solution to this conflict.

%============================
\section{Weak Quantum Discord}
\label{Sec:WQD}

The reduction axiom prescribes that after a measurement of an operator $B=\sum_bbB_b$ is performed and an outcome $b$ is obtained, the preparation $\rho$ on $\cal{H_A\otimes H_B}$ collapses according to the map
\be 
C_{b|B}(\rho):=\frac{(\mathbbm{1}\otimes B_b)\rho\,(\mathbbm{1}\otimes B_b)}{\text{Tr}\left[(\mathbbm{1}\otimes B_b)\,\rho\,(\mathbbm{1}\otimes B_b)\right]}=\rho_{\cal{A}|b}\otimes B_b.
\label{C}
\ee 
Given a projective measurement of this type, the observer is granted with full information about the reduced state ($\rho_{\cal{B}}=B_b$) of the system. We now employ a map that allows us to effectively interpolate between weak and projective measurements:
\be
C_{b|B}^{\epsilon}(\rho):=(1-\epsilon)\,\rho+\epsilon\,C_{b|B}(\rho),
\label{Ce}
\ee 
with $\epsilon\in(0,1)$. Clearly, $C_{b|B}^{\epsilon}$ represents a strong projective measurement for $\epsilon\to 1$ and no measurement at all for $\epsilon\to~0$. For $0<\epsilon<1$ the map implies a small change in the preparation $\rho$, thus suitably simulating the notion of a weak measurement. Several properties can be derived for the map \eqref{Ce}, in particular that $\lim_{n\to \infty}[C_{b|B}^{\epsilon}]^n=C_{b|B}$. If no information is revealed about the outcome $b$, then our prediction for the post-measurement state is the averaging
\be 
M_B^{\epsilon}(\rho):=\sum_bp_bC_{b|B}^{\epsilon}(\rho)=(1-\epsilon)\rho+\epsilon\,\Phi_B(\rho).
\label{Me}
\ee 
This map, which we refer to as {\em monitoring}, continuously connects a regime of no intervention, $M_B^{\epsilon\to 0}(\rho)=\rho$, with the one of a projective unrevealed measurement, $M_B^{\epsilon\to 1}(\rho)=\Phi_B(\rho)$. In between these extrema we have a weak unrevealed measurement of intensity $0<\epsilon<1$. It is worth noticing that this map respects the non-signaling principle ($\text{Tr}_{\cal{B}}[M_B^{\epsilon}(\rho)]=\text{Tr}_{\cal{B}}[\rho]$) and correctly implements the fact that infinitely many weak measurements are equivalent to projective measurements ($\lim_{n\to \infty}[M_B^{\epsilon}]^n=\Phi_B$). For a thorough discussion about the maps \eqref{C}-\eqref{Me} we refer the reader to Ref. \cite{dieguez18}, where these maps have been introduced.

Let us now compute the QD using the Singh and Pati procedure, which consists of replacing the original form \eqref{QD} with \eqref{SQD}. In terms of our weak-measurement map \eqref{Ce} we have
\be 
\mathfrak{D}_{\cal{B}}^{\epsilon}(\rho)=\min_B\sum_bp_bS(C_{b|B}^{\epsilon}(\rho))+S(\rho_{\cal{B}})-S(\rho).
\label{frakDe}
\ee 
It follows from the concavity and the additivity of the von Neumann entropy that
\beq 
\mathfrak{D}_{\cal{B}}^{\epsilon}(\rho)&>& \min_B\sum_bp_b\Big[(1-\epsilon)S(\rho)+\epsilon S(\rho_{\cal{A}|b}) \Big]+S(\rho_{\cal{B}})-S(\rho) \nonumber \\
&=&(1-\epsilon)S(\rho)+\epsilon \min_B\sum_bp_bS(\rho_{\cal{A}|b})+S(\rho_{\cal{B}})-S(\rho)+\epsilon\Big[S(\rho_{\cal{B}})-S(\rho_{\cal{B}})\Big] \nonumber \\
&=&\epsilon\Big[\min_B\sum_bp_bS(\rho_{\cal{A}|b})+S(\rho_{\cal{B}})-S(\rho)\Big]+(1-\epsilon)S(\rho_{\cal{B}}) \nonumber \\ &=&\epsilon D_{\cal{B}}(\rho)+(1-\epsilon)S(\rho_{\cal{B}}).
\eeq 
Since $\mathfrak{D}_{\cal{B}}^{\epsilon\to 0}(\rho)>S(\rho_{\cal{B}})$, the drawback of Singh and Pati's approach persists, that is, the definition \eqref{frakDe} is not able as well to predict, for any $\rho$, zero distance in the limit of no disturbance. 

Now we change the strategy. Given that the forms \eqref{QD} and \eqref{QDRS} are mathematically equivalent upon the use of projective measurements, one might think at a first sight that there is no reason {\em a priori} for one to prefer one of them when weak measurements are used instead. However, we should realize that a weak measurement does not provide a precise outcome on which we could apply the {\em conditioning}, so that the meaning of the form \eqref{QD}, which is based on the conditional entropy, becomes unclear in this case. We then take the form \eqref{QDRS} as the primitive notion of QD. In terms of the monitoring \eqref{Me}, this allows us to introduce the {\em weak quantum discord} (WQD):
\be 
\cal{D_B}^{\epsilon}(\rho):=\min_B\Big[I(\rho)-I(M_B^{\epsilon}(\rho)) \Big]\qquad (0<\epsilon<1),
\label{De}
\ee 
which clearly reduces to QD as $\epsilon\to 1$. Most importantly, this form trivially implements the feature we have been looking for, namely, $\cal{D_B}^{\epsilon\to 0}(\rho)=0\,\,(\forall \rho)$. We now prove a result that precisely defines the sense in which the quantum discord quantifier $\cal{D_B}^{\epsilon}$ can be termed genuinely {\em weak}.
\begin{theorem}
For any density operator $\rho$ on $\cal{H_A\otimes H_B}$ and $\epsilon$ real such that $\epsilon\in(0,1)$, the weak quantum discord \eqref{De} is never greater than the quantum discord \eqref{QDRS}, that is, $\cal{D_B}^{\epsilon}(\rho)\leqslant D_{\cal{B}}(\rho)$. The equality holds for quantum-classical states of the form $\Phi_B(\rho)=\rho$, in which case $\cal{D_B}^{\epsilon}=D_{\cal{B}}=0$.
\end{theorem}

\noindent Proof.---Consider an instance in which a system $\cal{AB}$ initially prepared in a density operator $\rho$ on $\cal{H_A\otimes H_B}$ ends up into $M_B^{\epsilon}(\rho)$ after the monitoring of a generic observable $B$ on $\cal{H_B}$. The Stinespring theorem \cite{nielsen00,dieguez18} ensures that this mapping can be cast in terms of an entangling dynamics $U(t)$ between $\cal{B}$ and some extra degree of freedom $\cal{X}$ initially prepared in a state $\ket{x_0}\bra{x_0}$, that is,
\be 
M_B^{\epsilon}(\rho)=\text{Tr}_{\cal{X}}\left[U(t)\,\rho\otimes\ket{x_0}\bra{x_0}\,U^{\dag}(t) \right]=\rho_{\cal{AB}}(t),
\label{rhot}
\ee 
with $U(t)$ acting on $\cal{H_B\otimes H_X}$. By direct application of the reduced trace we obtain $\text{Tr}_{\cal{A}}M_B^{\epsilon}(\rho)=M_B^{\epsilon}(\rho_{\cal{B}})=\rho_{\cal{B}}(t)$. In addition, $\rho_{\cal{BX}}(t)=U(t)\,\rho_{\cal{B}}\otimes\ket{x_0}\bra{x_0}\,U^{\dag}(t)$ and $\rho_{\cal{A}}(t)=\rho_{\cal{A}}$. The unitary invariance of the von Neumann entropy allows us to write $S(\rho_{\cal{ABX}}(t))=S(\rho)$ and $S(\rho_{\cal{BX}}(t))=S(\rho_{\cal{B}})$. From the strong subadditivity of the von Neumann entropy, $S(\rho_{\cal{ABX}}(t))+S(\rho_{\cal{B}}(t))\leqslant S(\rho_{\cal{AB}}(t))+S(\rho_{\cal{BX}}(t))$ \cite{nielsen00}, one then obtains
\be 
S(\rho)+S(M_B^{\epsilon}(\rho _{\cal{B}}))\leqslant S(M_B^{\epsilon}(\rho))+S(\rho_{\cal{B}}). 
\ee 
Since $M_B^{\epsilon}(\rho_{\cal{A}})=\rho_{\cal{A}}$, it immediately follows from the definition of mutual information that
\be 
I(\rho)\geqslant I(M_B^{\epsilon}(\rho)).
\label{monotI}
\ee 
This is a statement of the monotonicity of the mutual information under unrevealed weak measurements---an expected result since monitoring, as defined by the com\-ple\-tely-positive trace preserving map \eqref{Me}, is, after all, a quantum operation \cite{dieguez18}. Also, this proves that the WQD is non-negative. Since $M_B^{\epsilon}\Phi_B(\rho)=\Phi_B(\rho)$, it can be directly checked that the equality holds for $\rho=\Phi_B(\rho)=\sum_bp_b\rho_{\cal{A}|b}\otimes B_b$, that is, when the preparation $\rho$ is a quantum-classical state (a state of reality for the observable $B$ \cite{bilobran15}). In this case, both the WQD and the QD vanish\footnote{Of course the WQD also vanishes for $\epsilon\to 0$, but this trivial limit is not included in the statement of the Theorem 1.}.

We now employ the property $[M_B^{\epsilon}]^n(\rho)=(1-\epsilon)^n\rho+[1-(1-\epsilon)^n]\Phi_B(\rho)$, which has been proved in Ref. \cite{dieguez18} for the map \eqref{Me} and from which we can directly show that $[M_B^{\epsilon}]^{n\to\infty}(\rho)=\Phi_B(\rho)$. Via successive application of the relation \eqref{monotI} we obtain
\be 
I(M_B^{\epsilon}(\rho))\geqslant I([M_B^{\epsilon}]^2(\rho))\geqslant \cdots\geqslant I(\Phi_B(\rho)).
\label{f1}
\ee 
It then follows that 
\beq 
\cal{D_B}^{\epsilon}(\rho)&=&\min_B\Big[I(\rho)-I(\Phi_B(\rho))+I(\Phi_B(\rho))-I(M_B^{\epsilon}(\rho))\Big] \leqslant  D_{\cal{B}}(\rho),
\label{f2}
\eeq  
which completes the proof.\hfill{$\blacksquare$}

To emphasize the issue around the SQD, a remark is in order. Let us consider an unrevealed weak-measurement map $\Phi_{\{P_{\pm}\}}(\rho)=\sum_sP_s\rho P_s$ composed of the dichotomic operators \eqref{Ppm}. Since $\Pi_{0(1)}=\mathbbm{1}-\Pi_{1(0)}$ one shows, by direct manipulation, that
\be 
\Phi_{\{P_{\pm}\}}(\rho)=\mathrm{sech}(x)\,\rho+\big[1-\mathrm{sech}(x)\big]\Phi_{\Pi}(\rho)=M_{\Pi}^{1-\mathrm{sech}(x)}(\rho),
\ee 
which holds for all $x$ and for the operator $\Pi=\sum_s\pi_s\Pi_s$. This shows that the dichotomic map $\Phi_{\{P_{\pm}\}}$ is a specialization of the map $M_B^{\epsilon}$ and, as such, definitively allows for a proper definition of WQD in the molds of the proposal \eqref{De}. Hence, the conditioning to undefined outcomes turns out to be the only conceptual difficulty associated to the SQD proposal.

%==========================
\subsection{Interpretation}
\label{meaning}

The fact that the distance-based formulation \eqref{De} leads to $\cal{D_B}^{\epsilon\to 0}(\rho)=0$ even for discordant states, that is, states for which $D_{\cal{B}}(\rho)>0$, raises the question as to whether the WQD can be viewed as a faithful quantifier of quantum correlations. We now point out the precise meaning that we propose to attach to the WQD. Let us consider the measure $\mathfrak{a}(\rho,\sigma)=I(\rho)-I(\sigma)$ for any $\rho$ and $\sigma$ on $\cal{H_A\otimes H_B}$. With that, QD can be written as 
\be 
D_{\cal{B}}(\rho)=\min_B\,\,\mathfrak{a}(\rho,\Phi_B(\rho))=\mathfrak{a}(\rho,\Phi_{B_1}(\rho)), 
\ee 
where $B_1$ is the observable that implements the minimization. Using the traditional interpretation of QD we then take $\mathfrak{a}(\rho,\Phi_{B_1}(\rho))$ as the amount of quantum correlations that can be associated to $\rho$ under local measurements of the optimal observable $B_1$. For the WQD we similarly write $\cal{D_B}^{\epsilon}(\rho)=\min_B\,\,\mathfrak{a}(\rho,M_B^{\epsilon}(\rho))=\mathfrak{a}(\rho,M_{B_{\epsilon}}^{\epsilon}(\rho))$, where $B_{\epsilon}$ denotes the ($\epsilon$-dependent) optimal observable. 

Now, consider the state $\tilde{\rho}_{\epsilon}=M_{B_{\epsilon}}^{\epsilon}(\rho)$, which refers to a scenario in which a preparation $\rho$ has undergone a monitoring of the observable $B_{\epsilon}$. Since $\Phi_BM_B^{\epsilon}(\rho)=\Phi_B(\rho)$ for all $B$, then $\Phi_{B_{\epsilon}}(\tilde{\rho})=\Phi_{B_{\epsilon}}(\rho)$. Via direct manipulations one can show that 
\be \label{int}
\cal{D_B}^{\epsilon}(\rho)=\mathfrak{a}(\rho,\Phi_{B_{\epsilon}}(\rho))-\mathfrak{a}(\tilde{\rho}_{\epsilon},\Phi_{B_{\epsilon}}(\tilde{\rho}_{\epsilon})),
\ee 
which settles the interpretation for the WQD. The first term on right-hand side of Eq. \eqref{int} refers to the amount of quantum correlations encoded in $\rho$, whereas the second encapsulates the amount of quantum correlations that persist after the monitoring of the optimal observable $B_{\epsilon}$. Thus, $\cal{D_B}^{\epsilon}(\rho)$ can be viewed as {\em the amount of quantum correlations that is removed from $\rho$ by local weak measurements of $B_{\epsilon}$}. In consonance with this interpretation, we see for $\epsilon\to 0$ that the above formula readily gives $\cal{D_B}^{\epsilon\to 0}(\rho)=0$, meaning that no quantum correlation is destroyed when no measurement is performed. On the other hand, for $\epsilon\to 1$, the second term of Eq. \eqref{int} vanishes and we find $\cal{D_B}^{\epsilon\to 1}(\rho)=\mathfrak{a}(\rho,\Phi_{B_1}(\rho))=D_{\cal{B}}(\rho)$, meaning that all quantum correlations can be destroyed via projective measurements of the optimal observable.

%===================
\subsection{Example}
\label{example}

\begin{figure}[htb]
\centerline{\includegraphics[scale=0.45]{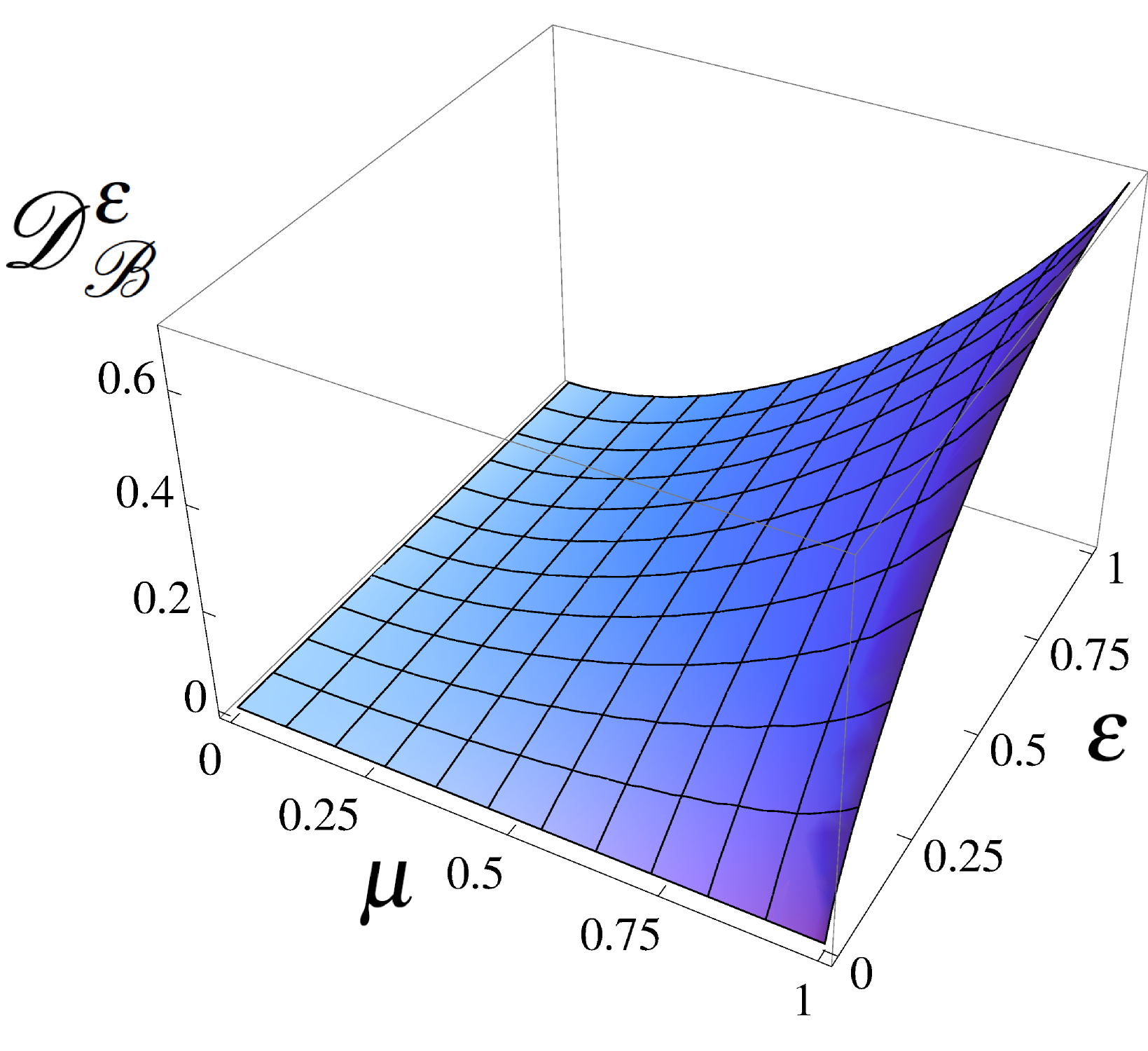}}
\caption{Weak quantum discord $\cal{D_B}^{\epsilon}(\rho^{\mu})$, as given by Eq. \eqref{Derhomu}, for the state $\rho^{\mu}$ as a function of $\mu$ and the strengh $\epsilon$ of the measurement. This is an illustration of Theorem 1, since $\cal{D_B}^\epsilon(\rho^{\mu})<\cal{D_B}^{\epsilon\to 1}(\rho^{\mu})=D_{\cal{B}}(\rho^{\mu})$ $\forall \epsilon\in(0,1)$. In particular $\cal{D_B}^{\epsilon\to 0}(\rho^{\mu})=0$.}
\label{fig1} 
\end{figure}

Let us consider the one-parameter state of two qubits:
\be
\rho^{\mu}=(1-\mu )\frac{\mathbbm{1}\otimes \mathbbm{1}}{4}+\mu\ket{s}\bra{s}, 
\ee
with the singlet state $\ket{s}=\frac{1}{\sqrt{2}}(\ket{01} -\ket{10})$. Noticing that $\rho^{\mu}_{\cal{B}}=M_B^{\epsilon}(\rho^{\mu}_{\cal{B}})=\mathbbm{1}/2$, one can use the definition of mutual information to reduce the WQD \eqref{De} to
\be 
\cal{D_B}^{\epsilon}(\rho^{\mu})=\min_B\Big[S(M_B^{\epsilon}(\rho^{\mu}))-S(\rho^{\mu})\Big].
\label{DeDS}
\ee 
The eigenvalues of $\rho^{\mu}$ are given by $\{\tfrac{1-\mu}{4},\tfrac{1-\mu}{4},\tfrac{1-\mu}{4},\tfrac{1+3\mu}{4}\}$. To compute the eigenvalues of $M_B^{\epsilon}(\rho^{\mu})$ we introduce the generic observable $B=\sum_{b=\pm}bB_b$ with projectors $B_{\pm}=\ket{\pm}\bra{\pm}$ such that $\ket{+} =\cos{\left(\tfrac{\theta}{2}\right)}\ket{0} +e^{i\phi}\sin{\left(\tfrac{\theta }{2}\right)}\ket{1}$ and $\ket{-}=-\sin{\left(\tfrac{\theta}{2}\right)}\ket{0} +e^{i\phi}\cos{\left(\tfrac{\theta}{2}\right)}\ket{1}$. Then we can directly compute $\Phi_B(\rho^{\mu})=\sum_s(\mathbbm{1}\otimes B_s)\rho^{\mu}(\mathbbm{1}\otimes B_s)$ and the weakly measured state $M_B^{\epsilon}(\rho^{\mu})=(1-\epsilon)\rho^{\mu}+\epsilon\Phi_B(\rho^{\mu})$, whose eigenvalues can be shown to be $\{\tfrac{1-\mu}{4},\tfrac{1-\mu}{4},\tfrac{1+3\mu-2\mu\epsilon}{4},\tfrac{1-\mu+2\mu\epsilon}{4}\}$. As a consequence of the rotational invariance of the singlet state, this  set has no information about the parameters $(\theta,\phi)$, which would be used for the minimization process. With the pertinent eigenvalues at hand, we can evaluate the entropies in Eq. \eqref{DeDS} and then finally write the WQD in compact form as
\be
\cal{D_B}^{\epsilon}(\rho^{\mu})=\tfrac{1}{4}\sum_{i=-1}^1\sum_{j=0}^1(-1)^j\lambda_{ij}\ln{\lambda_{ij}}, \qquad \lambda_{ij}=1+\mu[1+2i(1-j\epsilon)].
\label{Derhomu}
\ee 
This function is plotted in Fig. \ref{fig1}, where we can see that it indeed has the behavior expected for a genuine WQD: it is always less than QD, as implied by Theorem 1, that is, $\cal{D_B}^\epsilon(\rho^{\mu})<\cal{D_B}^{\epsilon\to 1}(\rho^{\mu})=D_{\cal{B}}(\rho^{\mu})$ $\forall \epsilon\in(0,1)$, and disappears for a vanishing monitoring [$\cal{D_B}^{\epsilon\to 0}(\rho^{\mu})=0$].

%%%%%%%%%%%%%%%%%%%%%%%%%
%%%%%%%%%%%%%%%%%%%%%%%%%

%=========================================
\section{Symmetrical Weak Quantum Discord}
\label{Sec:SyWQD}

In Ref. \cite{rulli11}, Rulli and Sarandy defined the {\em symmetrical quantum discord} (SyQD)
\be 
D(\rho)=\min_{A,B}\Big[I(\rho)-I(\Phi_A\Phi_B(\rho))\Big],
\label{SyQD}
\ee 
for observables $A=\sum_aaA_a$ and $B=\sum_bbB_b$ acting on $\cal{H_A}$ and $\cal{H_B}$, respectively,  projectors $\{A_a,B_b\}$, and the map $\Phi_A(\rho)=\sum_a(A_a\otimes\mathbbm{1})\rho(A_a\otimes\mathbbm{1})$ in analogy with the map \eqref{PhiB}. This quantifier is ``symmetrical'' in that both parties of the system are measured. Inspired by this definition, we introduce
\be 
\cal{D}^{(\epsilon',\epsilon)}(\rho):=\min_{A,B}\Big[I(\rho)-I(M_A^{\epsilon'}M_B^{\epsilon}(\rho))\Big]\qquad (0<\{\epsilon',\epsilon\}<1),
\label{SyWQD}
\ee 
as a quantifier of {\em symmetrical weak quantum discord} (SyWQD). The monitoring of the observable $A$ is given by $M_A^{\epsilon'}=(1-\epsilon')\rho+\epsilon'\Phi_A(\rho)$ in analogy with the monitoring \eqref{Me}. From the monotonicity of the mutual information [see relation \eqref{monotI}], it follows that SyWQD is a non-negative quantity. Also, $\cal{D}^{(\epsilon',\epsilon)}(\rho)$ will be zero only if $\rho=\Phi_A\Phi_B(\rho)=\sum_{a,b}p_{a,b}A_a\otimes B_b$, that is, for a state with no SyQD (a classical-classical state)\footnote{Of course, the SyWQD will also vanishes for $(\epsilon',\epsilon)\to(0,0)$.}. In light of the interpretation proposed in Sec. \ref{meaning}, we claim that the SyWQD should be viewed as a measure of the amount of quantum correlations that is removed form the state $\rho$ by local weak measurements. This position is supported by the fact that $\cal{D}^{(\epsilon',\epsilon)\to(0,0)}(\rho)=0$ and $\cal{D}^{(\epsilon',\epsilon)\to(1,1)}(\rho)=D(\rho)$.

%===================================
\subsection{Hierarchy and Ordering}

Interestingly, by use of the monotonicity of the mutual information \eqref{monotI} and the procedures employed to prove Theorem 1 [see formulas \eqref{monotI}-\eqref{f2}] we can make some statements about ordering and hierarchy of discord measures. By adding and subtracting $I(\Phi_{B}(\rho))$ in Eq. \eqref{SyQD}, we can prove that $D\geqslant D_{\cal{B}}$ (of course, it also holds that $D\geqslant D_{\cal{A}}$). This relation defines a {\em hierarchy}, in the same sense as discussed in Refs. \cite{costa16,gomes18}. This means that here we have a specific direction of implication between two quantities that are conceptually different. To appreciate this point, consider a quantum-classical state $\rho=\Phi_B(\rho)$. While for this state the QD is zero, the SyQD may not be. That is, the SyQD quantifies correlations that cannot be destroyed solely by measurements of $B$. It follows that not all symmetrically discordant state (those with $D>0$) are discordant states (those for which $D_{\cal{B}}>0$), while the converse is true. In other words, discordant states form a subset of symmetrically discordant states, so that the detection of QD for a given state immediately implies the existence of SyQD for this state. By introducing $I(M_{B}^{\epsilon}(\rho))$ in the definition \eqref{SyWQD} one finds $\cal{D}^{(\epsilon',\epsilon)}\geqslant \cal{D_B}^{\epsilon}$ (and, analogously,  $\cal{D}^{(\epsilon',\epsilon)}\geqslant \cal{D}_{\cal{A}}^{\epsilon}$), which shows that an equivalent hierarchy applies for the corresponding weak quantifiers.

On the other hand, let us similarly introduce $I(\Phi_A\Phi_B(\rho))$ in the Eq. \eqref{SyWQD} and use the monotonicity again. In this case we arrive at $\cal{D}^{(\epsilon',\epsilon)}\leqslant D$, which is a mere statement of {\em ordering}. That is, a simple comparison relation between two quantities that identify the same class of quantum correlations, namely, those that are destroyed by local monitorings in the sites $\cal{A}$ and $\cal{B}$. In fact, notice that both the SyWQD and SyQD vanish only for classical-classical states of the form $\rho=\Phi_A\Phi_B(\rho)$, thus meaning that set of symmetrically discordant states (those for which $D>0$) and the set of symmetrically weakly discordant states (those with $\cal{D}^{(\epsilon',\epsilon)}>0$) are one and the same. Similar conclusions apply for the quantifiers appearing in the Theorem~1.

%====================
\section{Conclusion}
\label{Sec:C}

We have shown that if we take a distance-based formulation as a primitive notion for quantum discord, as pondered in Refs. \cite{rulli11,modi10}, then no surprise is found when replacing projective measurements with weak ones. In particular, no ``super'' quantum discord emerges. Rather, we find a quantifier---the weak quantum discord---that interpolates between the regime of ``no quantum correlations destroyed'' (when no measurement is conducted, that is, $\epsilon\to 0$) and the regime of ``all quantum correlations destroyed'' (when a projective measurement is conducted, that is, $\epsilon\to 1$), in which case the quantum discord is recovered. This allows us to interpret the weak quantum discord as a measure of the amount of quantum correlations that is removed via local weak measurements. In addition, we have shown how to properly define a symmetrical weak quantum discord and briefly discussed notions of hierarchy and ordering among various discord-like quantifiers.

An important question is left open. It is well known that quantum discord reduces to the entanglement entropy for pure states, that is, $D_{\cal{B}}(\ket{\psi})=S(\rho_{\cal{A(B)}})$, for reduced states $\rho_{\cal{A(B)}}=\text{Tr}_{\cal{B(A)}}\ket{\psi}\bra{\psi}$. This can be proved by taking $B$ as the operator whose projectors $\ket{b_i}\bra{b_i}$ define the Schmidt decomposition $\ket{\psi}=\sum_i\sqrt{\lambda_i}\ket{a_i}\ket{b_i}$. By virtue of Theorem 1 we can directly conclude that $\cal{D_B}^{\epsilon}(\ket{\psi})\leqslant S(\rho_{\cal{A(B)}})$, but this does not allows us to claim that $\cal{D_B}^{\epsilon}(\ket{\psi})$ is an entanglement monotone. To this end one should be able to prove that $\cal{D_B}^{\epsilon}(\ket{\psi})$ does not increase on average under local measurements and classical communication \cite{vidal00}. Actually, to be fair, the very weak quantum discord (along with all its related quantifiers) is to be submitted to some reliability criteria \cite{brodutch12}, at least the non-debatable ones, before we can definitively assert that it is a genuine measure of quantum correlations.

%========================================
\begin{acknowledgements}
P.R.D. and R.M.A. respectively acknowledge financial support from the Brazilian Agencies CAPES and CNPq. This work was partially supported by the National Institute for Science and Technology of Quantum Information (INCT-IQ/CNPq, Brazil).
\end{acknowledgements}

%=========================================

\end{document}